\documentclass{article}
\usepackage{spconf,amsmath,graphicx}
\usepackage{multirow}
\usepackage{url}
\usepackage{makecell}

\title{IMPROVED SPEECH PRE-TRAINING WITH Supervision-ENHANCED ACOUSTIC UNIT}
\name{Pengcheng Li$^{1, 2}$, Genshun Wan$^{1, 2, 3}$,Fenglin Ding$^{1, 2}$, Hang Chen$^{3}$,Jianqing Gao$^{2}$, Jia Pan$^{1, 2}$, Cong Liu$^{1, 2}$}
\address{$^1$Shanghai AI Laboratory, China \\
	$^2$iFLYTEK Research, iFLYTEK Co. Ltd., China \\
	$^3$University of Science and Technology of China, China \\
	\small{\texttt{\{flding2, gswan, pcli2, jqgao, jiapan,congliu2\}@iflytek.com, ch199703@mail.ustc.edu.cn}}}

\begin{document}
\ninept
\maketitle
\begin{abstract}
Speech pre-training has shown great success in learning useful and general latent representations from large-scale unlabeled data. Based on a well-designed self-supervised learning pattern, pre-trained models can be used to serve lots of downstream speech tasks such as automatic speech recognition. In order to take full advantage of the labed data in low resource task, we  present an improved pre-training method by introducing a supervision-enhanced acoustic unit (SEAU) pattern to intensify the expression of comtext information and ruduce the training cost. Encoder representations extracted from the SEAU pattern are used to generate more representative target units for HuBERT pre-training process. The proposed method, named SeHuBERT, achieves a relative
word error rate reductions of 10.5\% and 4.9\% comared with the standard HuBERT on Turkmen speech recognition task with 500 hours and 100 hours fine-tuning data respectively. Extended to more languages and more data, SeHuBERT can aslo achieve a relative word error rate reductions of  approximately 10\%  at half of the training cost compared with HuBERT.

\end{abstract}
\section{Introduction}
\label{sec:intro}
The research on automatic speech recognition (ASR) system has seen great achievement in the recent decades~\cite{1-end2endSynnaeve2019,2-ContextnetHan2020,3-ConformerGulati2020,4-SpeechstewWilliam2021,5-AchievingXiong2016}. However, ASR has suffered from the reliance on human annotations, which is too costly or impractical for the popularization and application of most languages around the world. Therefore, there has been a lot of interest in how to better use unlabeled speech data~\cite{6-AdversarialLiu2018,7-SemiBaskar2019,8-SemiHsu2020},especially in low resource languages.

The most widely used method in this study can be classified into two categories: classical self-training and self-supervised learning (SSL).Classical self-training method achieves releative  considerable improvement  ~\cite{2-ContextnetHan2020,9-2019Lessons,10-2019Self,11-Iterative2020Xu,12-2020Improved} by pseudo-labeling unannotated audio data and then retraining the final system with the additional labeled data. Self-supervised learning typically utilizes speech pre-training to learn better representations. Based on a pre-trained model, the recognition system only needs to fine-tune on a small amount of labeled data ~\cite{13-2020wav2vec} and can achieve a competitive performance compared with the supervised models trained with quantities of labeled data. Moreover, speech pre-training is easier to operate compared with the complex cycle training process and careful selection strategy of pseudo labels in classical self-training method. Therefore, speech pre-training has received a lot of attention and achieved great success ~\cite{13-2020wav2vec,14-Iterative2020Xu,15-2019wav2vec,16-2020vq,17-2021W2v,18-2021HuBERT,19-2021WavLM,20-2022data2vec}.

As one of the effective pre-training methods in recent studies, Hidden-Unit BERT (HuBERT) ~\cite{18-2021HuBERT} benefits from an offline clustering step to provide target labels for a BERT-like prediction loss ~\cite{21-2018BERT}. However, like most other pre-training methods, HuBERT targets at learning a model that is able to produce the universal speech representations, which can be used for multiple downstream tasks. Instead, the labeled data used in the downstream tasks may not provide more help to the pre-training process. In terms of ASR task, We argue that the performance could be further improved if the network focus on learning content information with some guidance information such as labeling information. Meanwhile, as the standard configuration option, multiple iteration pattern of the pre-training model may increase the difficulty of promotion. Obviously, better initial target labels from supervised model can reduce the training cost and improve the specific representation while most methods do not fully exploit these advantages in low resource application tasks.

To this end, we use a supervised model trained with the labeled data which can be used in the downstream task, to guide the generation of  target labels for HuBERT pre-training. We call the newly designed paradigm supervision-enhanced hidden unit BERT (SeHuBERT) which is better suited to the low resource application task. The experiments are mainly conducted on a Turkmen speech recognition task. Compared with the standard HuBERT, the results show that the proposed SeHuBERT method achieves a significant improvement for different scales of data in just one iteration. Extended to more languages and more data, SeHuBERT also has stable performance improvement at half of the training cost.

The rest of this paper is organized as follows. Section 2 gives a brief description of the related work. The paradigm of the proposed SeHuBERT pre-training method is introduced in Section 3. Section 4 mainly shows our experimental setup and other details, including the experiment results and ablations. Finally, the discussion and conclusion is presented in Section 5.

\begin{figure*}[t]
    \centering
    \includegraphics[width=0.8\textwidth]{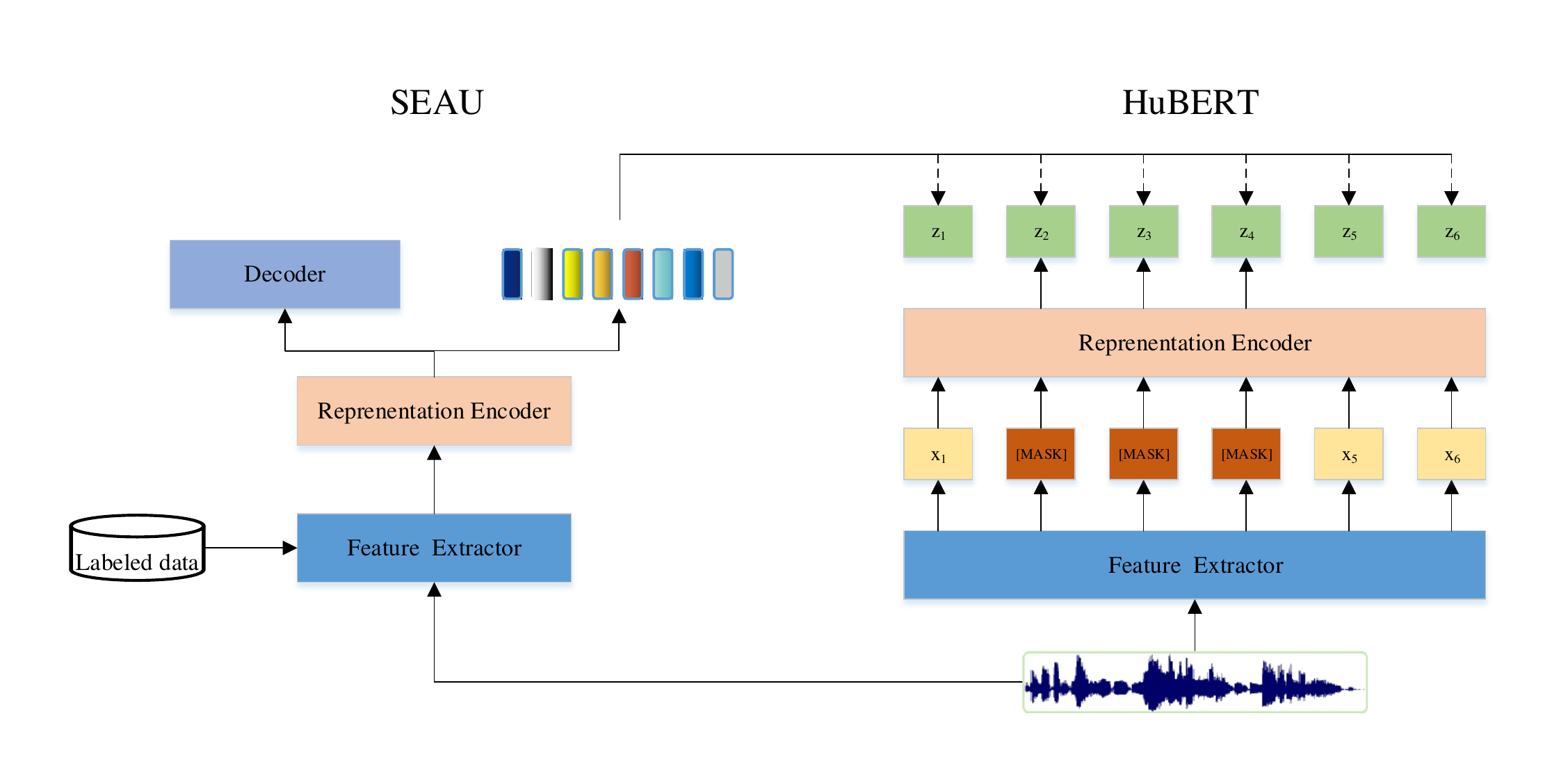}
    \caption{Illustration of SeHuBERT.}
    \label{fig:figure1}
\end{figure*}

\section{Related Works}
\label{sec:related}

HuBERT ~\cite{18-2021HuBERT} is a typical approach for self-supervised speech representation learning, which utilizes an offline clustering step to generate aligned target labels for a BERT-like pre-training. Researchers argue that wav2vec 2.0 only explores quantizing the waveform encoder output, which may not be the best feature for quantization due to the limited capacity of the convolutional feature extractor. Different from wav2vec 2.0, HuBERT benefits from the offline clustering step due to the consistency of the targets, not just their correctness. The generated target labels are also called acoustic units or target codes. What is more, the prediction loss is only applied over the masked regions, forcing the model to learn a combined acoustic and language model over the continuous inputs.

HuBERT pre-training adopts an iterative re-clustering and re-training process: For the first iteration, the discrete targets are obtained by clustering on the MFCC features extracted from the raw audio data. For the second iteration, the training targets are updated by clustering the hidden representations of the HuBERT model after the first iteration. According to the authors' studies, acoustic units generated from the middle to top layers have much higher mutual information score than other layers, indicating that they are better at phonetic information learning.

\section{Proposed Method}
\label{sec:proposed}
The standard HuBERT paradigm provides more universal speech representations suited for different speech tasks. In specific low-resource recognition tasks, a small amount of labeled data is still readily available. Utilizing the prior information from a few labeled data in the pre-training stage is of great interest and worth exploring. Combining some guidance information from downstream labeled data and pre-training strategy, we could take full advantage of self-training advantages to improve the performance of representation at a lower training cost.

Starting with the enhancement of these guidance information, we propose an improved speech pre-training paradigm named supervise-enhanced acoustic unit HuBERT (SeHuBERT) for these low resource application task.The architecture of SeHuBERT is illustrated as \figurename~\ref{fig:figure1}. We firstly train a supervised model with a certain amount of labeled speech data from the downstream tasks. Then the unsupervised speech-only data are fed to the supervised model encoder to generate the continuous deep representations. After that, simple discrete latent variable models such as k-means are used to generate discrete target units with the deep representations for HuBERT pre-training. Both supervised and discretized models together constitute the SeHuBERT paradigm.

\subsection{Supervision-enhanced acoustic unit pattern (SEAU)}
In standard HuBERT pre-training process, the discrete targets are assigned by clustering on the MFCC features for the first iteration. However, MFCC features are artificial designed, which may have certain limitations in modeling acoustic contextual information. Furthermore, the model quality of the first iteration directly influences the effect of the second iteration. Therefore, instead of generating acoustic units from MFCC features, we use the deep representations extracted from a supervised model to strengthen the content presentation and reduce the training cost.

We use the attention-based encoder-decoder (ED) model as the baseline architecture in this work. For SEAU pattern, firstly, a speech recognition model is trained with the labeled data in downstream task to minimize the cross entropy loss as follows:
\begin{equation}
	L(\theta;\boldsymbol{X},\boldsymbol{Y})=-\sum_{i=1}^Nlogp(\boldsymbol{Y}|\boldsymbol{X},\theta)
\end{equation}
Where $\boldsymbol{X}$ is the sequence of input speech frames, $\boldsymbol{Y}$ is the corresponding sequence of labels, which contains N tokens. $\theta$  are the model parameters to be optimized.

Once speech recognition model has converged, it can be used to extract the deep context representations. To get the frame-level aligned target units, we choose the output of  deep encoder layer from the supervised model as clustering features. Then, for the speech-only pre-training data, the deep representations can be obtained from the forward inference through the chosen  encoder layers. Finally, the continuous deep representations are discretized by a clustering model to obtain the pre-training targets. 
The SEAU pattern is denoted with $\boldsymbol{Z}=h(\boldsymbol{Encoder}(\boldsymbol{X}))= [\boldsymbol{z}_1,\boldsymbol{z}_2,\dots,\boldsymbol{z}_T]$, where $\boldsymbol{z}_t \in [\boldsymbol{C}]$ is a class categorical variable, and  $h$  is a clustering model such as k-means.

The deep features of speech recognition network contain higher-dimensional acoustic information, and thus are more robust to speaker variations and noise interferences compared to the MFCC features. More importantly, since the model is trained on labeled data, the deep features contain rich accurate contextual information. Therefore, the final clustering acoustic units force the pre-training model concentrate on learning content information, which is helpful to the ASR task.

Obviously, we can also use the forced alignment from a hybrid model ~\cite{22-2012Deep} or a connectionist temporal classification (CTC) ~\cite{23-2006Connectionist} model as the target units. Therefore, the proposed SEAU is generic for different acoustic models. Moreover, the process of clustering is similar to the decision tree clustering and state binding in hybrid models, which can merge similar tokens together and get more robust target units.

\subsection{HuBERT Pre-training}
After the supervision-enhanced acoustic units are obtained in an offline clustering step, we use them to pre-train the HuBERT model. The backbone of HuBERT in our work includes a feature extractor and a context representation encoder. During pre-training process, the feature extractor module takes the  frame-level speech  as the input to generate a higher dimensional features. Then some frames are randomly selected and furtuer masked. The corrupted features $\boldsymbol{\tilde{X}}$ are then fed into the encoder and  transformed into the hidden states $\boldsymbol{H}$ at the top layer. The network is optimized to predict the discrete target units $\boldsymbol{Z}$. The distribution over target units is parameterized as:
\begin{equation}
	p(c|\boldsymbol{\tilde{X}},t)=\frac{exp(sim(\boldsymbol{W}h_t),e_c/\tau)}{\sum_{c^{'}=1}^Cexp(sim(\boldsymbol{W}h_t),e_c^{'}/\tau)}
\end{equation}
Where $\boldsymbol{W}$ is a projection matrix, $h_t$ is the hidden state at frame $t$, $e_c$ is the embedding for unit c, $sim(a, b)$ computes the cosine similarity between two vectors and $\tau = 0.1$ scales the logit. A key ingredient of HuBERT is the prediction loss is only applied over the masked regions, forcing the model to learn  high-level representations of unmasked inputs to infer the targets of masked ones correctly.

In this work, we replace transformer block ~\cite{22-2012Deep} in standard HuBERT with conformer ~\cite{4-SpeechstewWilliam2021} beacuse conformer has presented higher effectiveness in ASR task. The major component of Conformer is a stack of conformer blocks, each of which is a series of multi-headed self-attention, depth-wise convolution and feed-forward layers. On the other hand, to improve computational efficiency, we use filter-bank features as input of network instead of raw waveforms. The feature extractor that acts as a convolutional sub-sampling block is replaced with two 2D-convolution layers, both with strides $(2, 2)$, resulting in a 4x reduction in the acoustic sequence length.

\subsection{Fine-tuning}
\label{subsec:teacher}
For simplicity, we keep the ASR network structure the same as supervised model in section 3.1 during fine-tuning stage. The difference is that the encoder is initialized with the pre-trained model. Simple yet effective, transformer decoder is kept. We insert a linear layer between the pre-trained HuBERT model and the decoder layer as the projection block. The whole ASR network is optimized to minimize the cross entropy loss with labeled data.

\section{Experiments}
\subsection{Datasets}
We mainly evaluate the proposed methods on a Turkmen speech recognition task with low resource. The entire dataset has only 500 hours of training data. We randomly select 100 hours as a low-resource subset. In addition, we use another dataset as out-domain data to evaluate the ability of SEAU to learn cross-lingual representations. The out-domain data contains 476 hours of Hungarian, a language quite distinct from Turkmen. We also randomly choose 100 hours as a subset. The test set has 1000 utterances of Turkmen, total in about 3 hours. All the speech files are sampled at 16K Hz with 16 bits.Furthermore, a Portuguese recognition task with 1600 hours training data and a French recognition task with 4300 hours training data are introduced to prove the practicability and generalization of SeHuBERT.

\subsection{Model Setups}
The main framework of the supervised model in both SEAU and fine-tuning stage are kept consistent for simplicity. The whole ASR encoder consists of a feature extractor and a representation encoder. Feature extractor has 2 convolutional layers with ?lter size (3, 3) and strides (2, 2). Representation encoder consists of 16 Conformer blocks, each with encoder embedding 512, hidden dimension 2048, 8 attention heads and convolution kernel size 15. The final projection layer dimension is 768. In total, the whole encoder including feature extractor has about 100 million trainable parameters. The decoder consists of 6 transformer decoder layers, each with decoder embedding 512, hidden dimension 2048, 8 attention heads. We take specaugment [28] with mask parameter (F=15, T=40) as a basic configuration. The inputs
of the model are the 40-dimensional log Mel-scale filter-bank
features. Modeling units are 15k BPE sub-words ~\cite{31-2015Neural} trained from the labeled data.

\subsection{Training Details}
We train a standard HuBERT model with our modified architecture as the pre-training baseline. We follow the configuration of the acoustic unit pattern and pre-training configures described in ~\cite{18-2021HuBERT}.
All subsequent experiments were performed by the fairseq toolkit ~\cite{29-2019fairseq} and run
on a server equipped with 4 v100 GPUs with adam optimizer ~\cite{30-2014Adam}.

\textbf{SEAU pattern}. For in-domain evaluation, we train the supervised model on 100 hours and 500 hours of Turkmen data respectively. While for out-domain evaluation, we train the supervised model on 100-hours subset of Hungarian data. For all the supervised models, the entire 500 hours of Turkmen speech are forwarded through the model encoder and produce deep representations. We run k-means clustering with 1000 clusters on the supervised deep representations for the HuBERT training. As a whole, We train the supervised model for 15 epochs which is a small price compared with the pre-training and fine-tuning process.

\textbf{Pre-training}. We train the SeHuBERT model for one iteration with 400k steps. Dropout 0.1 was used in the conformer and at the output of the feature extractor. Layers were dropped at a rate of 0.2. We optimize with Adam, warming up the learning rate for the first 8000 updates to a peak of 0.0005, and then linearly decay it. We also regularize the model by applying a L2 penalty to the activations of the final layer of the feature extractor and scale down the gradients for the extractor by a factor of 10.

\textbf{Fine-tuning}. The fine-tuning step is similar to the supervised training except that the model encoder is initialized from the pre-trained SeHuBERT model.  The learning rate warms up linearly from 0 to the peak learning rate 0.0007 for the first 8000 training steps, decays to the 15-th epoch and then is halved for each epoch from the 16-th epoch. During fine-tuning, the feature extractor parameters were fixed.

\subsection{Results and Discussion}
\tablename~\ref{tab:tab1} shows the word error rate (WER) of the no pre-training, HuBERT and SeHuBERT model on Turkmen test set with various amount of fine-tuning data while the training data used in SEAU and HuBERT are all fixed 500 hours. Iter-1 and iter-2 stands for the first and  second iteration of HuBERT respectively. It can be seen that HuBERT significantly outperforms the model without pre-training, achieving a relative 20.3\% and 61.9\% WER reduction respectively, which shows the importance of pre-training to the performance of low-resource speech recognition task. Furthermore, the proposed SeHuBERT method achieves a relative 4.9\% WER reduction compared with the final result of standard HuBERT on a 500 hours data set and a relative 10.5\% WER reduction on a 100 hours data set. Along with the significant recognition performance improvement, the new paradigm cut their training cost in half.

\begin{table}[t]
	\caption{WER (\%) results of  the proposed method on Turkmen }
	\label{tab:tab1}
	\centering
	\setlength{\tabcolsep}{7mm}{
	\begin{tabular}{c | c  c }
		\hline
		\hline
		\multirow{2}{*}{Method} & \multicolumn{2}{c}{fine-tuning data} \\
		
		& 100h & 500h \\
		\hline
		No pre-training & 53.63 & 16.20 \\
		HuBERT-iter1 & 35.86 & 16.29  \\
		HuBERT-iter2 & 20.41 & 12.91 \\
		SeHuBERT & 18.26 & 12.28 \\
		\hline
		\hline
	\end{tabular}
}
\end{table}

We also presented the results on  Portuguese  and French task. The results shown in \tablename~\ref{tab:tab4} are consistent with the results on Turkmen task. The proposed SeHuBERT method achieves a relative 8.7\% WER on Portuguese task and a relative 10.5\% WER reduction over the standard HuBERT model. In terms of both effectiveness and efficiency, the proposed SeHuBERT may be more advisable.

 \begin{table}[h]
 	\caption{WER (\%) results of  SeHuBERT on different languages  }
 	\label{tab:tab4}
 	\centering
 	\setlength{\tabcolsep}{7mm}{
 		\begin{tabular}{c | c  c }
 			\hline
 			\hline
 			\multirow{1}{*}{Method} &  Portuguese &  French \\
 			\hline

 			HuBERT-iter2 & 7.14 & 6.29 \\
 			SeHuBERT & 6.52 & 5.63 \\
 			\hline
 			\hline
 		\end{tabular}
 	}
 \end{table}

\subsection{Ablation Studies}
\textbf{Amount of supervised data in SEAU Pattern:} Intuitively, the quality of the supervised model in SEAU affects the effectiveness of the target units used for pre-training. Therefore, we first explore the impact of the amount of supervised data in SEAU, which is closely related to the performance of the supervised model. We use the 100h subset of Turkmen to train the supervised model in SEAU and compare it with the entire 500h set. The results are shown in \tablename~\ref{tab:tab2}. Overall, the performance of SEAU with 500h data is better than that of 100h data. But the gap narrows as the model is fine-tuned with more labeled data. The quality of the targets generated from the SEAU pattern with 100h labeled data are worse than that with 500h labeled data. Therefore, with a well-trained supervised model for SEAU, HUBERT can achieve good performance with only one iteration.

\begin{table}[h]
	\caption{WER (\%) results of different amount of supervised data used in SEAU pattern}
	\label{tab:tab2}
	\centering
	\setlength{\tabcolsep}{7mm}{
		\begin{tabular}{c | c  c }
			\hline
			\hline
			\multirow{2}{*}{\makecell[c]{fine-tuning data \\ based on SeHuBERT} } & \multicolumn{2}{c}{data used in SEAU} \\
			
			& 100h & 500h \\
			\hline
			100h  & 20.45 & 18.26 \\
			500h & 13.02 & 12.28  \\
			\hline
			\hline
		\end{tabular}
	}
\end{table}

\textbf{Source of supervised data in SEAU Pattern:}We then investigate the effect of the data from different source  for the supervised model in SEAU pattern. We use a 100h subset of Hungarian to train the supervised model to generate the target units. It is expected that the SEAU can learn cross-lingual speech information. The Hungarian subset is denoted as out-domain, while the corresponding 100h Turkmen subset is denoted as in-domain. As shown in \tablename~\ref{tab:tab3}, out-domain case achieves a competitive performance compared with the in-domain source on 100h fine-tuning data. Though on 500h fine-tuned data, out-domain performs worse than in-domain, it is comparable to the standard HuBERT model shown in table \tablename~\ref{tab:tab1}.

\begin{table}[t]
	\caption{WER (\%) results of different source of supervised data used in SEAU pattern}
	\label{tab:tab3}
	\centering
	\setlength{\tabcolsep}{4mm}{
		\begin{tabular}{c | c  c }
			\hline
			\hline
			\multirow{2}{*}{\makecell[c]{fine-tuning data \\ based on SeHuBERT} } & \multicolumn{2}{c}{domains used in SEAU/100h } \\
			
			& In-domain & Out-domain \\
			\hline
			100h  & 20.45 & 20.21 \\
			500h & 13.02 & 13.65  \\
			\hline
			\hline
		\end{tabular}
	}
\end{table}

\textbf{Clustering Number in SEAU Pattern:} The clustering number is another factor that influence the pre-training performance. We compare the impact of different clustering number in the case of different amount of supervised data used in SEAU pattern. The results are summarized in \tablename~\ref{tab:tab4}, $C$ refers to the clustering number. We find that a larger number of clusters achieves better performance when using 100h supervised data in SEAU pattern. However, for 500h in SEAU, the conclusion is the opposite. It is speculated that the result is in a normal fluctuation range.

\begin{table}[h]
	\caption{WER (\%) results of different cluster numbers used in SEAU pattern}
	\label{tab:tab4}
	\centering
	\setlength{\tabcolsep}{4mm}{
		\begin{tabular}{c | c  c }
			\hline
			\hline
			\multirow{2}{*}{\makecell[c]{Clustering number of SEAU} } & \multicolumn{2}{c}{data used in SEAU } \\
			
			& 100h & 500h \\
			\hline
			C500  & 20.62 & 17.94 \\
			C1000 & 20.45 & 18.26  \\
			\hline
			\hline
		\end{tabular}
	}
\end{table}

\section{Conclusion}
In this work, we propose a novel speech pre-training method named SeHuBERT, in which a supervised model is utilized to guide the generation of the acoustic units  for HuBERT pre-training. The idea of the proposed method is to force the model concentrate on learning content information during pre-training. Experimental results show that based on a supervised model without detailed design, the proposed SeHuBERT achieves relative
word error rate reductions of 10.5\% and 4.9\% comared with the standard HuBERT on 500h and 100h Turkmen fine-tuning data.  Extended to more languages and more data, SeHuBERT also has about relative 10\% performance improvement at half of the training cost.



\bibliographystyle{IEEEbib}
\bibliography{strings,refs}

\begin{thebibliography}{10}

\bibitem{1-end2endSynnaeve2019}
Gabriel Synnaeve, Qiantong Xu, Jacob Kahn, Tatiana Likhomanenko, Edouard Grave,
  Vineel Pratap, Anuroop Sriram, Vitaliy Liptchinsky, and Ronan Collobert,
\newblock ``End-to-end asr: from supervised to semi-supervised learning with
  modern architectures,''
\newblock {\em arXiv preprint arXiv:1911.08460}, 2019.

\bibitem{2-ContextnetHan2020}
Wei Han, Zhengdong Zhang, Yu~Zhang, Jiahui Yu, Chung-Cheng Chiu, James Qin,
  Anmol Gulati, Ruoming Pang, and Yonghui Wu,
\newblock ``Contextnet: Improving convolutional neural networks for automatic
  speech recognition with global context,''
\newblock {\em arXiv preprint arXiv:2005.03191}, 2020.

\bibitem{3-ConformerGulati2020}
Anmol Gulati, James Qin, Chung-Cheng Chiu, Niki Parmar, Yu~Zhang, Jiahui Yu,
  Wei Han, Shibo Wang, Zhengdong Zhang, Yonghui Wu, et~al.,
\newblock ``Conformer: Convolution-augmented transformer for speech
  recognition,''
\newblock {\em arXiv preprint arXiv:2005.08100}, 2020.

\bibitem{4-SpeechstewWilliam2021}
William Chan, Daniel Park, Chris Lee, Yu~Zhang, Quoc Le, and Mohammad Norouzi,
\newblock ``Speechstew: Simply mix all available speech recognition data to
  train one large neural network,''
\newblock {\em arXiv preprint arXiv:2104.02133}, 2021.

\bibitem{5-AchievingXiong2016}
Wayne Xiong, Jasha Droppo, Xuedong Huang, Frank Seide, Mike Seltzer, Andreas
  Stolcke, Dong Yu, and Geoffrey Zweig,
\newblock ``Achieving human parity in conversational speech recognition,''
\newblock {\em arXiv preprint arXiv:1610.05256}, 2016.

\bibitem{6-AdversarialLiu2018}
Alexander~H Liu, Hung-yi Lee, and Lin-shan Lee,
\newblock ``Adversarial training of end-to-end speech recognition using a
  criticizing language model,''
\newblock pp. 6176--6180, 2019.

\bibitem{7-SemiBaskar2019}
Murali~Karthick Baskar, Shinji Watanabe, Ramon Astudillo, Takaaki Hori,
  Luk{\'a}{\v{s}} Burget, and Jan {\v{C}}ernock{\`y},
\newblock ``Semi-supervised sequence-to-sequence asr using unpaired speech and
  text,''
\newblock {\em arXiv preprint arXiv:1905.01152}, 2019.

\bibitem{8-SemiHsu2020}
Wei-Ning Hsu, Ann Lee, Gabriel Synnaeve, and Awni Hannun,
\newblock ``Semi-supervised speech recognition via local prior matching,''
\newblock {\em arXiv preprint arXiv:2002.10336}, 2020.

\bibitem{9-2019Lessons}
Sree Hari~Krishnan Parthasarathi and Nikko Strom,
\newblock ``Lessons from building acoustic models with a million hours of
  speech,''
\newblock in {\em ICASSP 2019-2019 IEEE International Conference on Acoustics,
  Speech and Signal Processing (ICASSP)}. IEEE, 2019, pp. 6670--6674.

\bibitem{10-2019Self}
Jacob Kahn, Ann Lee, and Awni Hannun,
\newblock ``Self-training for end-to-end speech recognition,''
\newblock in {\em ICASSP 2020-2020 IEEE International Conference on Acoustics,
  Speech and Signal Processing (ICASSP)}. IEEE, 2020, pp. 7084--7088.

\bibitem{11-Iterative2020Xu}
Qiantong Xu, Tatiana Likhomanenko, Jacob Kahn, Awni Hannun, Gabriel Synnaeve,
  and Ronan Collobert,
\newblock ``Iterative pseudo-labeling for speech recognition,''
\newblock {\em arXiv preprint arXiv:2005.09267}, 2020.

\bibitem{12-2020Improved}
Daniel~S Park, Yu~Zhang, Ye~Jia, Wei Han, Chung-Cheng Chiu, Bo~Li, Yonghui Wu,
  and Quoc~V Le,
\newblock ``Improved noisy student training for automatic speech recognition,''
\newblock {\em arXiv preprint arXiv:2005.09629}, 2020.

\bibitem{13-2020wav2vec}
Alexei Baevski, Yuhao Zhou, Abdelrahman Mohamed, and Michael Auli,
\newblock ``wav2vec 2.0: A framework for self-supervised learning of speech
  representations,''
\newblock {\em Advances in Neural Information Processing Systems}, vol. 33, pp.
  12449--12460, 2020.

\bibitem{14-Iterative2020Xu}
Aaron van~den Oord, Yazhe Li, and Oriol Vinyals,
\newblock ``Representation learning with contrastive predictive coding,''
\newblock {\em arXiv preprint arXiv:1807.03748}, 2018.

\bibitem{15-2019wav2vec}
Steffen Schneider, Alexei Baevski, Ronan Collobert, and Michael Auli,
\newblock ``wav2vec: Unsupervised pre-training for speech recognition,''
\newblock {\em arXiv preprint arXiv:1904.05862}, 2019.

\bibitem{16-2020vq}
Alexei Baevski, Steffen Schneider, and Michael Auli,
\newblock ``vq-wav2vec: Self-supervised learning of discrete speech
  representations,''
\newblock 2019.

\bibitem{17-2021W2v}
Yu-An Chung, Yu~Zhang, Wei Han, Chung-Cheng Chiu, James Qin, Ruoming Pang, and
  Yonghui Wu,
\newblock ``W2v-bert: Combining contrastive learning and masked language
  modeling for self-supervised speech pre-training,''
\newblock pp. 244--250, 2021.

\bibitem{18-2021HuBERT}
Wei-Ning Hsu, Benjamin Bolte, Yao-Hung~Hubert Tsai, Kushal Lakhotia, Ruslan
  Salakhutdinov, and Abdelrahman Mohamed,
\newblock ``Hubert: Self-supervised speech representation learning by masked
  prediction of hidden units,''
\newblock {\em IEEE/ACM Transactions on Audio, Speech, and Language
  Processing}, vol. 29, pp. 3451--3460, 2021.

\bibitem{19-2021WavLM}
Sanyuan Chen, Chengyi Wang, Zhengyang Chen, Yu~Wu, Shujie Liu, Zhuo Chen, Jinyu
  Li, Naoyuki Kanda, Takuya Yoshioka, Xiong Xiao, et~al.,
\newblock ``Wavlm: Large-scale self-supervised pre-training for full stack
  speech processing,''
\newblock {\em IEEE Journal of Selected Topics in Signal Processing}, 2022.

\bibitem{20-2022data2vec}
Alexei Baevski, Wei-Ning Hsu, Qiantong Xu, Arun Babu, Jiatao Gu, and Michael
  Auli,
\newblock ``Data2vec: A general framework for self-supervised learning in
  speech, vision and language,''
\newblock {\em arXiv preprint arXiv:2202.03555}, 2022.

\bibitem{21-2018BERT}
Jacob Devlin, Ming-Wei Chang, Kenton Lee, and Kristina Toutanova,
\newblock ``Bert: Pre-training of deep bidirectional transformers for language
  understanding,''
\newblock {\em arXiv preprint arXiv:1810.04805}, 2018.

\bibitem{22-2012Deep}
Geoffrey Hinton, Li~Deng, Dong Yu, George~E Dahl, Abdel-rahman Mohamed, Navdeep
  Jaitly, Andrew Senior, Vincent Vanhoucke, Patrick Nguyen, Tara~N Sainath,
  et~al.,
\newblock ``Deep neural networks for acoustic modeling in speech recognition:
  The shared views of four research groups,''
\newblock {\em IEEE Signal processing magazine}, vol. 29, no. 6, pp. 82--97,
  2012.

\bibitem{23-2006Connectionist}
Alex Graves, Santiago Fern{\'a}ndez, Faustino Gomez, and J{\"u}rgen
  Schmidhuber,
\newblock ``Connectionist temporal classification: labelling unsegmented
  sequence data with recurrent neural networks,''
\newblock pp. 369--376, 2006.

\bibitem{31-2015Neural}
Rico Sennrich, Barry Haddow, and Alexandra Birch,
\newblock ``Neural machine translation of rare words with subword units,''
\newblock {\em arXiv preprint arXiv:1508.07909}, 2015.

\bibitem{29-2019fairseq}
Myle Ott, Sergey Edunov, Alexei Baevski, Angela Fan, Sam Gross, Nathan Ng,
  David Grangier, and Michael Auli,
\newblock ``fairseq: A fast, extensible toolkit for sequence modeling,''
\newblock {\em arXiv preprint arXiv:1904.01038}, 2019.

\bibitem{30-2014Adam}
Diederik~P Kingma and Jimmy Ba,
\newblock ``Adam: A method for stochastic optimization,''
\newblock {\em arXiv preprint arXiv:1412.6980}, 2014.

\end{thebibliography}

\end{document}